\documentclass[review]{elsarticle}

\usepackage{amssymb}
\usepackage{textgreek}
\usepackage{caption}
\usepackage{subcaption}
\usepackage{lipsum}  
\usepackage{xcolor, soul}
\sethlcolor{yellow}
\usepackage{hyperref}

\date{April 19, 2022}

\journal{Nucl. Instrum. Methods Phys. Res. A}

\begin{document}

\begin{frontmatter}

\title{The Sparse Readout RIGEL Application Specific Integrated Circuit for Pixel Silicon Drift Detectors in Soft X-Ray Imaging Space Applications}

\author[PoliMi]{Massimo~Gandola}

\author[UniPv,INFNP]{Marco~Grassi}

\author[PoliMi,INFN]{Filippo~Mele}

\author[PoliMi,INFN]{Irisa~Dedolli}

\author[UniPv,INFNP]{Piero~Malcovati}

\author[PoliMi,INFN]{Giuseppe~Bertuccio}

\affiliation[PoliMi]{organization={Politecnico di Milano - Department of Electronics, Information and Bioengineering},%
            city={Como},
            country={Italy}}
            
\affiliation[UniPv]{organization={University of Pavia - Department of Electrical Computer and Biomedical Engineering},%
            city={Pavia},
            country={Italy}}

\affiliation[INFNP]{organization={Italian National Institute of Nuclear Physics - INFN sez. Pavia},%
            city={Pavia},
            country={Italy}}     
            
\affiliation[INFN]{organization={Italian National Institute of Nuclear Physics - INFN sez. Milano},%
            city={Milan},
            country={Italy}}

\begin{abstract}
An Application Specific Integrated Circuit (ASIC), called RIGEL, designed for the sparse readout of a Silicon Pixel Drift Detector (PixDD) for space applications is presented. The low leakage current (less than 1 pA at +20 °C) and anode capacitance (less than 40 fF) of each pixel (300 μm $\times$ 300 μm) of the detector, combined with a low-noise electronics readout, allow to reach a high spectroscopic resolution performance even at room temperature. The RIGEL ASIC front-end architecture is composed by a 2-D matrix of 128 readout pixel cells (RPCs), arranged to host, in a 300 μm-sided square area, a central octagonal pad (for the PixDD anode bump-bonding), and the full-analog processing chain, providing a full-shaped and stretched signal. In the chip periphery, the back-end electronics features 16 integrated 10–bits Wilkinson ADCs, the configuration register and a trigger management circuit. The characterization of a single RPC has been carried out whose features are: eight selectable peaking times from 0.5 μs to 5 μs, an input charge range equivalent to 30 keV, and a power consumption of less than 550 μW per channel. The RPC has been tested also with a 4$\times$4 prototype PixDD and 167 eV Full Width at Half Maximum (FWHM) at the 5.9 keV line of $^{55}$Fe at 0°C and 1.8 μs of peaking time has been measured.
\end{abstract}


\end{frontmatter}

\section{Introduction}
\label{sec:introduction}

Astrophysical compact objects are unique, natural laboratories to study the physics in the presence of extreme gravitational and magnetic fields. Understanding the way matter and radiation behave in the close surrounding of neutron stars and black holes yields critical insight into several key areas of research in astrophysics. The X-ray radiation emitted by matter under such conditions encodes the signatures of the fundamental physics processes which are among the main scientific drivers of the ADAM (Advanced Detector for X-Ray Astronomy Missions) and eXTP (enhanced X-ray Timing and Polarization) mission concepts \cite{zhang2017extp}. Several imaging detectors have been developed for high resolution X-ray spectroscopy such as pixel detectors \cite{carini2009tests}, Charge Coupled Devices (CCDs) \cite{meidinger2010ccd} and sensors based on Depleted p-channel Field Effect Transistors (DePFET) \cite{lutz2007depfet}. However, these solutions are not optimal for sparse readout applications due to their relatively long acquisition dead-times, susceptibility to pile-up events, and necessity of cooling the detector to achieve a good energy resolution. In the international technological context of radiation instrumentation for astrophysics experiments, this technological gap is thus observed in the capability to realize pixel silicon detectors with fast response, asynchronously detecting and triggering on the individual X-ray photons (not integrated imaging as in the case of CCDs or DePFETs), while still offering Fano-limited energy resolution and soft X-ray response (down to $\sim$0.5 keV). The ability to work at room temperature, is also key to largely reducing costs and complexity of the relevant space projects and it, virtually, open to other possible applications of the proposed detector,  for example in new generation radiation instrumentation for advanced light sources facilities \cite{rachevski2019xafs, bufon2018new, porro2021minisdd}.

\subsection{The Pixel Silicon Drift Detector (PixDD)}

A novel detection system, based on a Pixel Silicon Drift Detector (called PixDD) and an application specific integrated circuit (ASIC), named RIGEL, with the capability of sparse readout, fast timing response and excellent noise performance even at room temperature, was developed within this project. The PixDD detector, developed by INFN-Trieste and Fondazione Bruno Kessler (FBK), is a 2-D matrix of square pixels, where each pixel is a small drift detector with a central circular anode surrounded by two cathodes, common to all the pixels, necessary to set the drift field. The advantage of the PixDD pixel relies in the low capacitance (esteemed below 40 fF), and low dark current at the anode electrode (around 1 pA at +20 °C). These two characteristics allow to operate at shorter pulse shaping times with lower noise and achieving a higher energy resolution even at room temperature (+20 °C) or moderate cooling  (0 °C). A first 4$\times$4 PixDD prototype with pixel dimension of 500 μm $\times$ 500 μm has been characterized with the SIRIO low-noise preamplifier \cite{mele2021sirio} to determine the ultimate energy resolution performance of the detector, achieving 130 eV FWHM of $^{55}$Fe 5.9 keV at room temperature. \cite{sammartini2020pixel, evangelista2018characterization}. The second PixDD prototype contains 128 pixels in a 16$\times$8 array at a 300 μm pitch. For the second prototype, a multi channel (16$\times$8) ASIC called RIGEL, has been designed and produced, to be bump-bonded with the detector anodes, allowing a high level of integration and a significant reduction of the stray capacitance at the PixDD-RIGEL interface.

\begin{table}[]
   \centering \caption{RIGEL ASIC specifications}
    \label{tab:asic_requirements}
   \vskip 0.2cm
   \scalebox{0.90}{
	    \begin{tabular}{|l|c|}
			\hline
	      \textbf{Parameter} &  \textbf{Value} \\
	      \hline
	      \hline
	     RPC area & 300 μm $\times$ 300 μm  \\
	      \hline
	     Maximum input charge & 8400\ e$^{-}$ (30 keV) \\
	    \hline
	    Peaking times (selectable) & 0.5, 1.0, 1.5, 1.8, 3.8, 4.0, 4.5, 5.0 \textmu s \\
	    \hline
	    Maximum input DC current & 20 pA \\
	    \hline
	    Input load capacitance & 10 $-$ 100 fF \\
	    \hline
	    Equivalent noise charge & 8 $e^{-}_{rms}$ at $I_{pixel}$ = 5 pA \\
        \hline
        Non-linearity error & $\leq$ 0.5 \% \\
        \hline
        Threshold level global & 0.5$-$3.0 keV (8 Bits) \\
        \hline
        Threshold level local & $\pm 30\%$ (4 Bits) \\
        \hline
        RPC power consumption & {$\leq$ 0.55 mW/Channel} \\ 
        \hline
	    \end{tabular}
	    }
 \end{table}
\section{The RIGEL ASIC}

\begin{figure}[!t]
\centering
\includegraphics[width=\columnwidth]{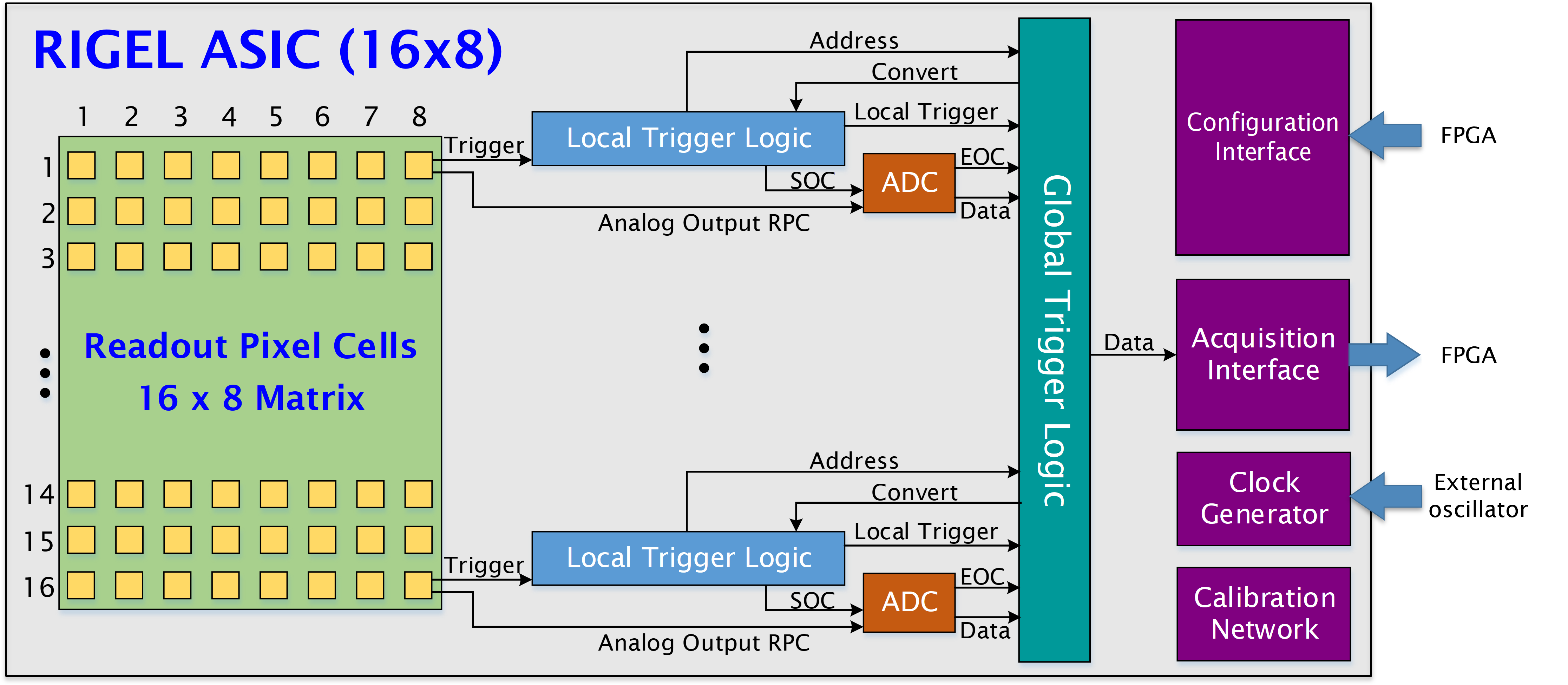}
\caption{Block diagram of the multi-channel RIGEL ASIC.}
\label{fig:rigel_bloc_diagram}
\end{figure}

The RIGEL ASIC is a multi channel, sparse read-out electronics, manufactured in a low-noise 0.35 \textmu m CMOS technology, designed to process signals delivered by the PixDD with the goal of performing spectral and timing studies of low-flux compact cosmic sources. The PixDD layout sets a series of challenges for the design of the readout electronics, which are summarized in Table~\ref{tab:asic_requirements}. A schematic representation of the RIGEL ASIC multi channel architecture is shown in Fig.~\ref{fig:rigel_bloc_diagram}: a front-end 2-D matrix of 128 readout pixel cells (RPCs), is arranged to be bump-bonded to the PixDD matrix using a flip-chip interconnection process \cite{caselle2016low}; in the 300 μm $\times$ 300 μm area of the RPC, a central input octagonal pad and the full-analog processing chain are hosted, providing a shaped and stretched signal. In the chip periphery, the back-end electronics for the triggering management, end-of-line signal processing, A/D conversion, configuration logic and communication interface with the external read-out board is implemented.
A detailed description of the front-end and back-end electronics is given in the following sections.  

\subsection{Readout Pixel Cells}
\begin{figure}[!b]
\centering
\includegraphics[width=\columnwidth]{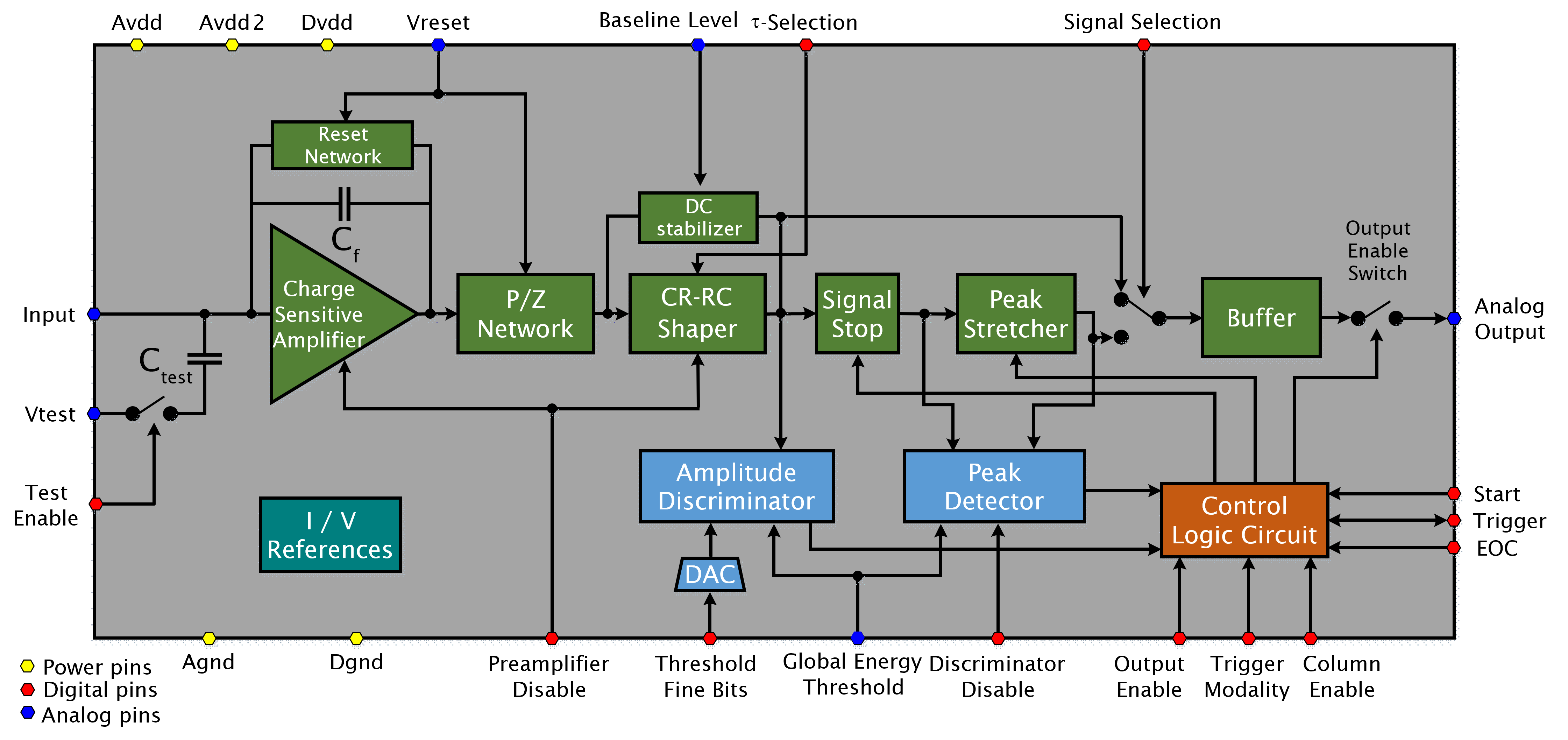}
\caption{Detailed architecture of the readoutp pixel cell (RPC) of the RIGEL ASIC.}
\label{fig:rigel_rpc_architecture}
\end{figure}

The block diagram of the single RPC is shown in Fig.~\ref{fig:rigel_rpc_architecture}. The first stage of the RPC is the charge sensitive amplifier (CSA), designed with specific criteria aimed to minimize the white and 1/f noise components considering the constraint on power consumption \cite{bertuccio2009noise}.
The CSA is based on a classical folded-cascode configuration using a p-MOSFET input transistor operating on a dedicated power supply, which allows some additional saving in the power consumption, and less disturbances from the other circuital blocks. An integrated test capacitance, excludable via a dedicated configuration bit, can be used for test or debug purposes.
This stage can also be disabled by setting a proper bit of the control logic register (pre\_disable) the feedback capacitance is short-circuited, so that the input charge is not integrated and the output is kept at the DC voltage of the input node. This option allows to safely stop the channel acquisition in case of problems, keeping the nominal bias condition at the pixel anode and, at the same time, protecting the input node from dangerous high voltage variation, which can occur in case of floating node conditions.

The CSA is followed by a high-linearity pole-zero compensation stage \cite{de1999cmos}, and a shaping current conveyor, responsible for the first CR shaping. The shaper implements eight user-defined shaping times from 0.5 to 5 \textmu s, selectable by switches driven by the back-end configuration logic.
The following stage is a first order RC semi-Gaussian shaper based on the R-lens block \cite{bertuccio1999r}, for the generation of the RC time constant, and a DC stabilizer for the stabilization of the DC output level. The selection of the shaping time is performed by choosing the capacitance value by means of three switches, driven by the same bits used in the first shaping stage. %

The shaped signal is compared by an amplitude discriminator with an internally programmable threshold that can be set from 500 eV to 3 keV in equivalent photon energy, through a chip-level coarse setting, and a local fine setting with a 4-bit DAC integrated in each RPC; when the amplitude discriminator toggles the peak-stretcher is activated, extending the duration of the peak amplitude in order to allow a proper conversion by the internal ADC. A peak-detector block recognizes when the peak amplitude of the acquired signal has been reached and generates a digital high signal indicating that the RPC output is ready to be converted by the ADC; the peak-detector allows also to have a consistent amplitude-independent timing information on the detected photons. The last stage is a voltage buffer used to decouple the RPC analog chain from the ADC (or other external loads which might be connected for debug purposes). The amplitude discriminator and the peak detector, combined with the Control Logic (CL), implement the pile-up rejection logic and internal reset signal. The total power consumption of the RPC is 540 \textmu W.

\begin{figure}[!b]
\centering
\includegraphics[width=0.8\columnwidth]{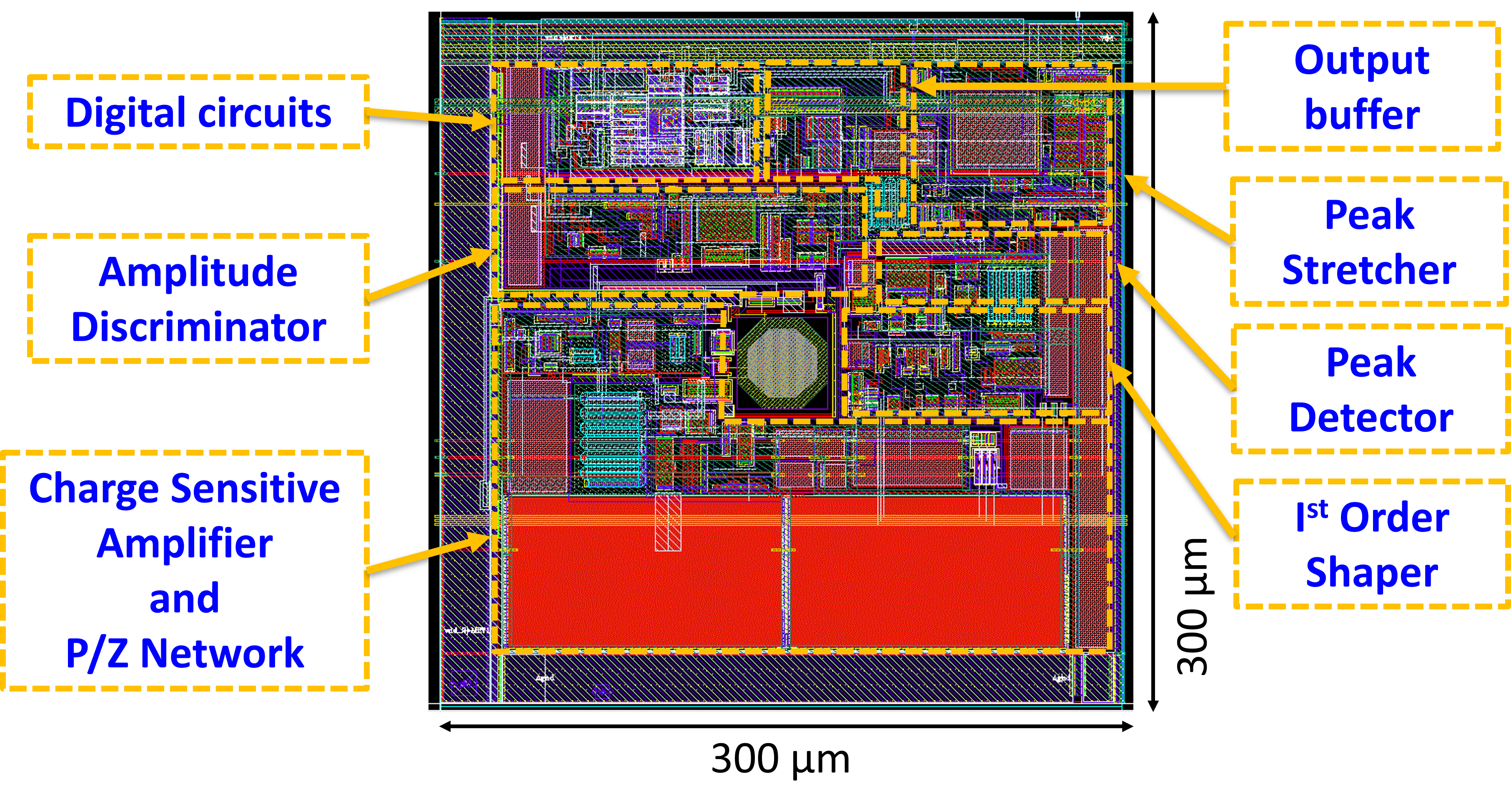}
\caption{Layout of a signal readout pixel cell (RPC) with highlighted positions of the functional blocks. The octagonal input pad, to be bump-bonded to the corresponding SDD pixel is visible in the central region.}
\label{fig:rigel_rpc_layout}
\end{figure}

The layout view of the single Readout Pixel Cell (RPC) is shown in Fig.~\ref{fig:rigel_rpc_layout}, where the octagonal input pad, to be bump-bonded to the corresponding SDD pixel is visible in the middle of the RPC. The internal connections between the components are made with polysilicon, metal 1 and metal 2 layers only. Metal 3 and metal 4 are used for global routing with external ADC, references, configuration circuits and power supplies. To avoid parasitic injection from ASIC to detector, in some areas of the RPC, a Metal 4 layer connected to ground has been used to shield the ASIC, especially in the latest processing stages where capacitive coupling between the full-dynamic range output signals and the close-by pixel could cause spurious charge injection and, eventually, a false detections in the front-end matrix. It is worthwhile to point out that, in addition to the much lower output capacitance, the drift structure of the pixels of PixDD has the advantage, over standard pn junction pixels, of the electrostatic shielding of the cathodes toward the sensitive anode in most of the pixel area.

\subsection{{Back-End Electronics}}

The main roles of the back-end electronics (BEE) are shown schematically in the block diagram of the ASIC in Fig.~\ref{fig:rigel_bloc_diagram}. They are the conversion of the analog signal coming from the RPC through a 10-bit Wilkinson ADC available for each row of the RPC matrix (16 ADCs in total) and the trigger and configuration management. Acquisition interface is also part of the back-end. The Local Trigger Logic (LTL) and Global Trigger Logic (GTL) circuits take care of managing the trigger signals coming from the RPC matrix and the A/D conversion of the analog output signals of the triggered RPCs in the correct sequence through suitable priority logic (to avoid
eventual conflicts). In particular, the trigger signals produced by the RPC matrix when an over threshold event occurs, are acquired and processed in order to produce a global trigger signal to be delivered to the Acquisition Interface. The data converted by the ADC and the address of the RPCs that have triggered, are stored in the GTL registers and, finally, sent to the Acquisition Interface, which is equipped with a 64 register FIFO memory. Energy values of all events are thus retrieved by the external payload manager, which manages the data flow with the ASIC. The RIGEL ASIC has three trigger functionalities: Normal Trigger Mode, External Trigger Mode, and OR Trigger Mode. When the Normal Trigger Mode is selected, each RPC can trigger independently and the trigger is generated by the pixel over threshold. On the other hand, in External Trigger Mode, a sample and hold signal freezes the analog output of all RPCs. The external trigger modality is useful, for instance, to cope with the charge sharing between adjacent pixels. Finally, in OR Trigger Mode the whole matrix is freezed and read-out on a single cell event (the matrix scan starts upon the first event over threshold). The high versatility of the RIGEL ASIC is guaranteed by a custom configuration interface and memory. Configuration data can be loaded with random access with a serial protocol and the stored information can also be read-out upon request for memory test purpose. The configuration is divided in three main parts: the common interface protocol management circuit, the global interface storage circuit, and the pixel configuration data storage array circuits. In fact, the configuration of each pixel (in terms for instance of pre-amplifier and discriminator enabling, shaping time selection and test enabling) are embedded in each row to minimize disturbances which may arise in case of unshielded or long wires from the configuration registers to the pixels gates.

\subsection{{The Wilkinson ADC}}

\begin{figure}[t]
\centering
\includegraphics[width=0.8\columnwidth]{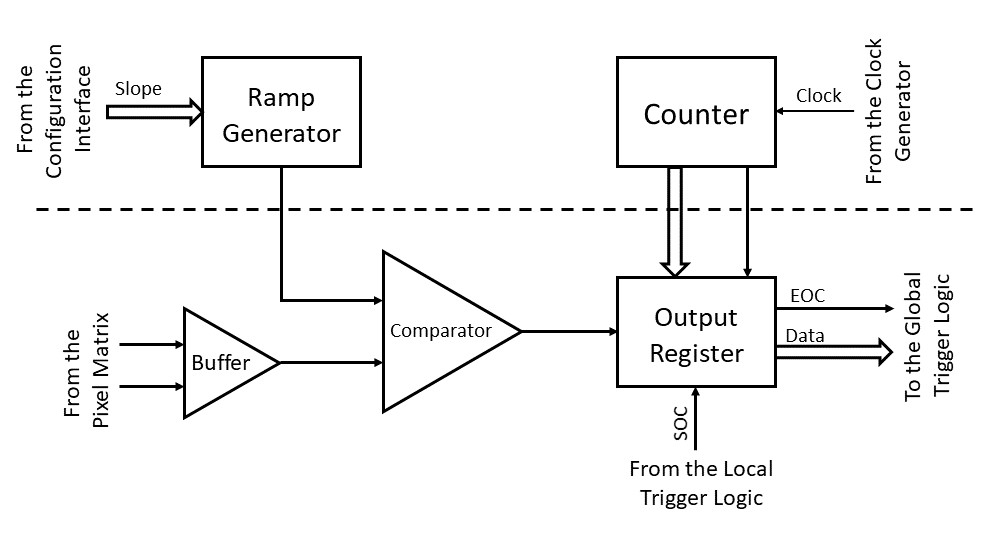}
\caption{Block diagram of the on-chip Wilkinson ADC.}
\label{fig:rigel_adc}
\end{figure}

The block diagram of the Wilkinson ADC is shown in Fig.~\ref{fig:rigel_adc}. The circuit consists of an Input Buffer, a Comparator, an Output Register, a Ramp Generator and a Counter. The Ramp Generator and the Counter are common to all of the ADCs, while there is an Input Buffer, an Output Register and a Comparator associated to each ADC input, which means for each row. Local Trigger Logic includes a counter section and a logic to drive all the ADCs. The fact that the ramp generator is common does not mean that the ADCs cannot work simultaneously, because sooner or later the common ramp will cross both the input of each ADC of interest and the full scale as explained below. The Comparators track the pixel output signal with the ramp produced by the Ramp Generator. When the Output Register is enabled by the Start of  Conversion (SOC) signal received from the Local Trigger Logic and the Comparator changes state because the ramp crossed the input signal, the content of the Counter is copied in the associated Output Register. When the Counter reaches full scale, the Output Register produces the End of Conversion (EOC) signal. With a clock frequency of 50 MHz (high speed operation), the Wilkinson ADC requires 20 µs of conversion time to achieve 10 bits of resolution. With a clock frequency of 10 MHz (standard speed operation), the conversion time is 100 µs. The slope of the Ramp Generator will be controlled with 7 bits provided by the Configuration Interface, in order to guarantee the required ADC gain in spite of any process variations. The Clock Generator receives an external differential sinusoidal signal at a frequency of 10 MHz up to 50 MHz and produces the 10–50 MHz clock required for operating the Wilkinson ADC. The differential sinusoidal input has been employed to minimize the cross-talk with the analog section of the ASIC. 

\subsection{{The Probe Network}}

The Probe Network allows the direct access to some RPC-internal analog signals for debug purposes. The Probe Network is constituted by blocks distributed inside the whole ASIC and some of them are shared between the probe network and other functional blocks, such as the calibration network and the output buffer interfaces used for the external acquisition and analysis of the analog output. The Probe Network is enabled by writing a “1” in the most significant bit (MSB) of a dedicated PROBE register, which is also used to specify the address (row/column) of the RPC under test, and the internal signal desired for probing (common to all the RPCs and valid only when the probe network is enabled). The enabling of the Probe Network automatically enables also the differential output buffer on the output interface. All RPCs except the probed one are disconnected from the analog bus to avoid cross-talk or short circuits on the analog bus of each row (the same bus that is connected to the input of the row ADC). On the probe net the operator may select among: shaped signal of a RPC, stretched signal of a RPC, or the trigger generated by the RPC. In a special probe mode, called probe-ADC, it is possible to force the input of the Wilkinson A/D converter from outside, disconnecting all RPCs from the analog buses. When the Probe Network is enabled the ADC will normally work under control from the Global Trigger Logic, but the converted output will have no relationship with the events causing the triggering, while in probe-ADC mode it is possible to choose a row of the matrix (i.e. a particular Wilkinson ADC) and to read-out its converted data applied by setup as well as having a look to the ramp which controls its timing.

\begin{figure}
     \centering
         \includegraphics[width=0.5\textwidth]{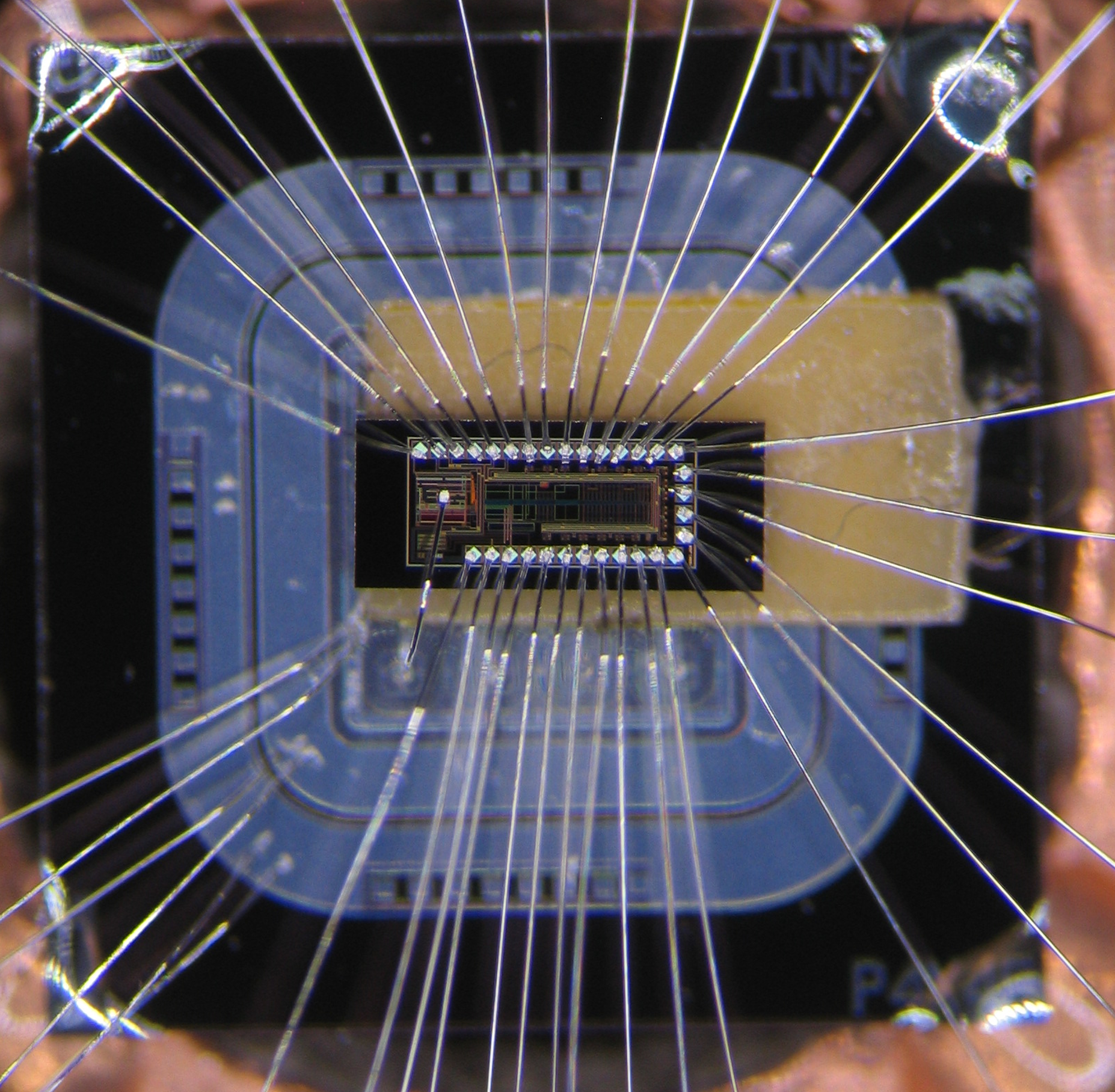}
        \caption{Micrograph of the single channel RIGEL RPC wire-bonded to a prototypical 4$\times$4 PixDD cell, used for the analog front-end performance characterization.}
        \label{fig:rigel_micrograph}
\end{figure}

\section{Experimental characterization}
A single RPC connected to one pixel of a 4 $\times$ 4 PixDD prototype detector with a pixel dimension of 500 \textmu m $\times$ 500 \textmu m has been characterized on a dedicated test board. In order to allow a fair characterization of the performances achievable with the front-end readout chain, the RPC was included on a dedicated production, excluding the back-end circuitry, and the Wilkinson ADC.

A micrograph of the single RPC used for the analog FEE performance characterization is shown in Fig.~\ref{fig:rigel_micrograph}.
The single RIGEL RPC coupled to the pixel detector has been characterized in terms of pulse shaping, pulse discrimination (noise threshold), and stretching functionality, linearity, dynamic range, and spectroscopic resolution. A pixel-level selection bit, allows to provide at the output of the RPC the shaped signal or the stretched signal. This function allows to test the channel using an external (off chip) commercial ADC. 

\subsection{Spectroscopic performance}
\begin{figure}[t]
\centering
\includegraphics[width=\columnwidth]{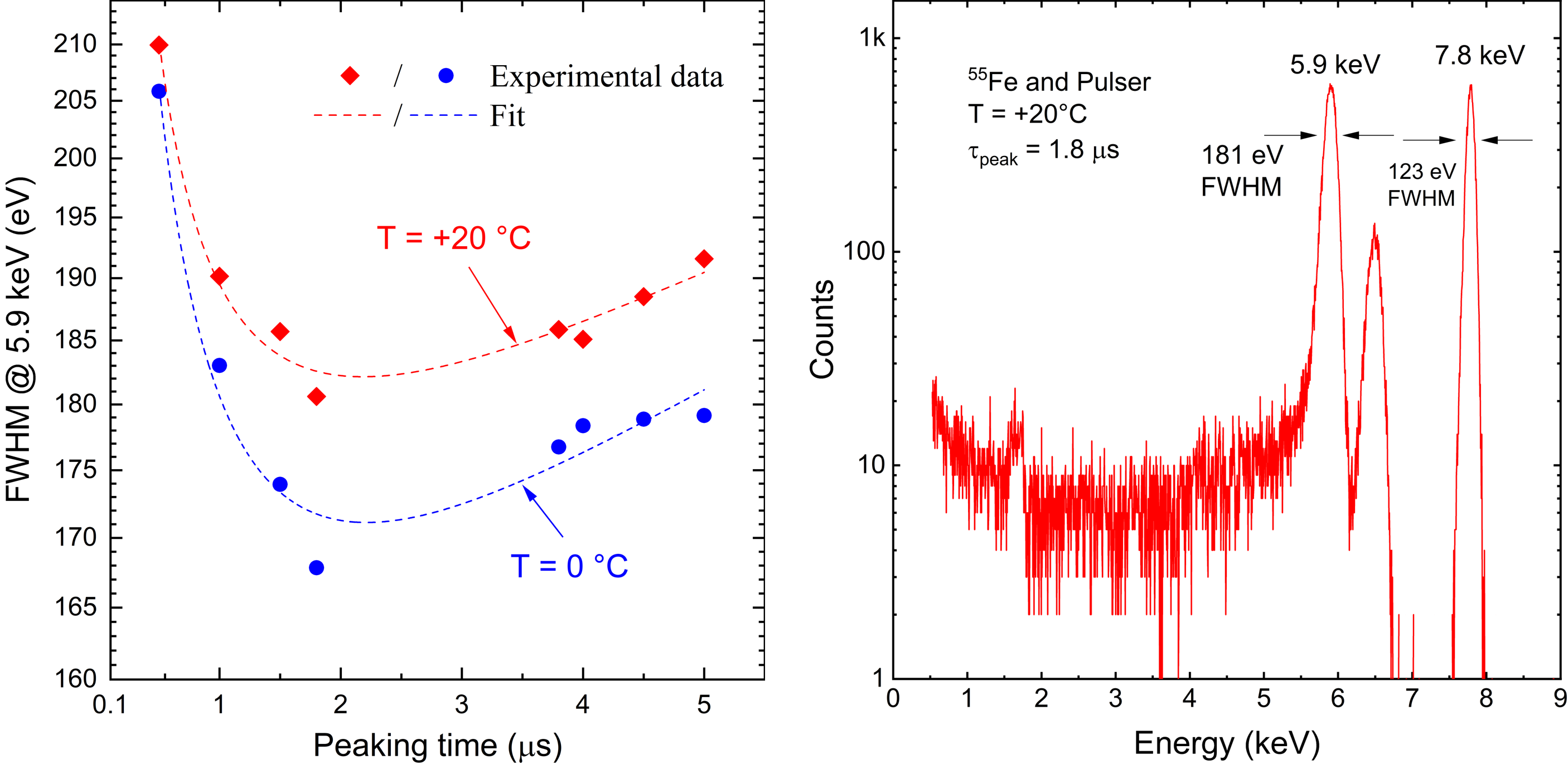}
\caption{On the left, the spectroscopic resolution in eV FWHM of the RIGEL ASIC for different shaping times, at 0 \textdegree C (blue curve) and +20 \textdegree C (red curve), as measured on the $^{55}$Fe 5.9 keV line. On the right, the energy spectrum acquired at +20\textdegree C with peaking time set to 1.8 \textmu s.}
\label{fig:combined_picture}
\end{figure}

The spectroscopic performance of the RIGEL RPC, coupled to the PixDD pixel, has been evaluated using an $^{55}$Fe calibration source to acquire different spectra according to the eight selectable peaking times of the detection system. The spectra have been acquired by sampling the shaped output of the single RPC with a commercial multi-channel analyzer (MCA). The full-width at half-maximum (FWHM) at the 5.9 keV line as a function of the peaking time is shown on the left of Fig.~\ref{fig:combined_picture} at room temperature (20 \textdegree C) and at moderate cooling (0 \textdegree C) using a climatic chamber with internal nitrogen flux to dry the test environment. The best result currently obtained was measured with 1.8 \textmu s of peaking time, where 167 eV FWHM have been measured at 0\textdegree C. The $^{55}$Fe spectrum, acquired at 20\textdegree C, with peaking time equal to 1.8 \textmu s is reported in the right part Fig.~\ref{fig:combined_picture}. The energy resolution, as measured on test pulse, is 123 eV FWHM, corresponding to an equivalent noise charge (ENC) of the system of 14.2 electrons r.m.s. (181 eV FWHM on the 5.9 keV Mn-K$\alpha$ line).
 Improvements to the spectral resolution are expected with the bump bonding.

\subsection{Linearity and energy dynamic range}
\begin{figure}[t]
\centering
\includegraphics[width=0.8\columnwidth]{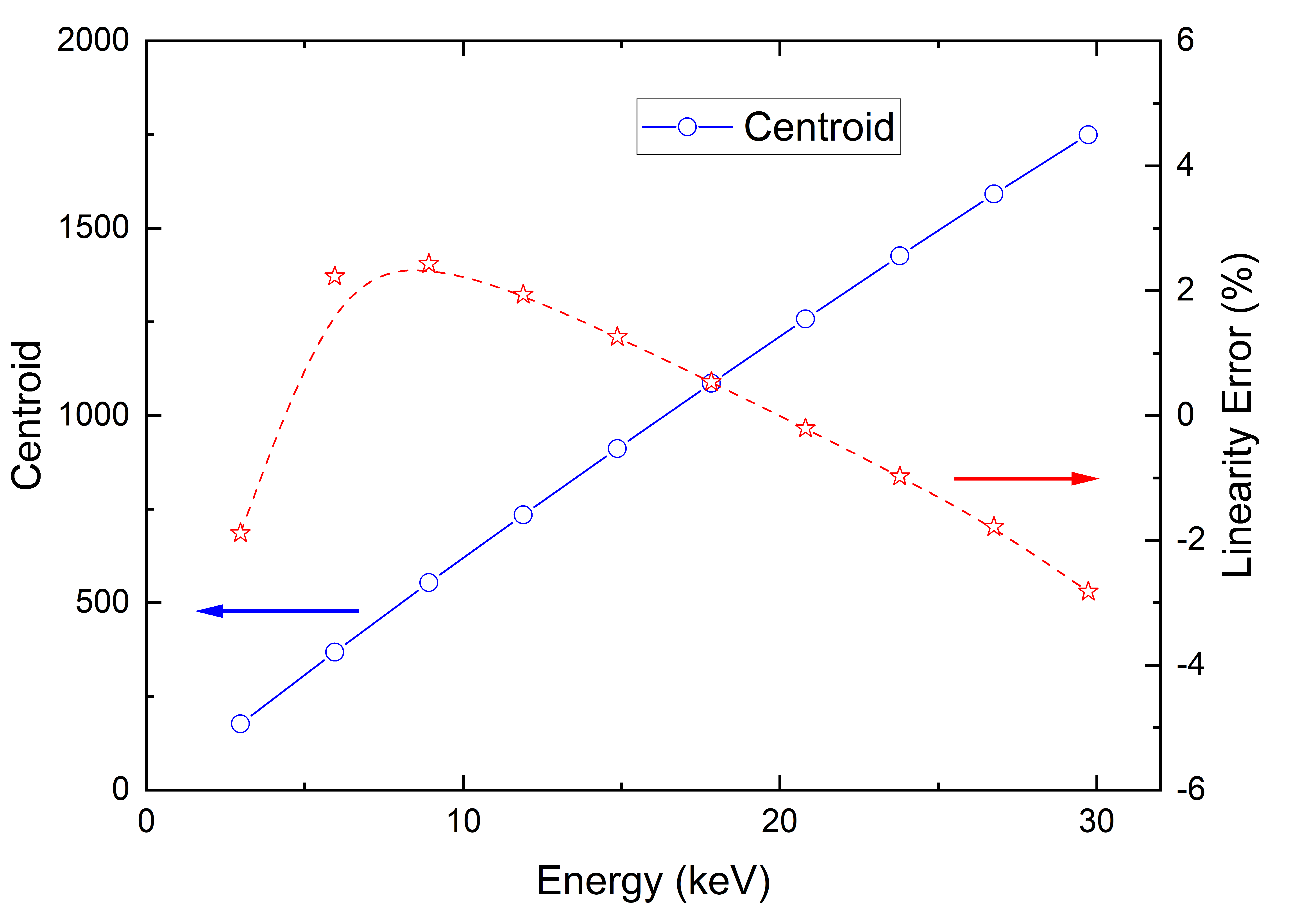}
\caption{Linearity error measured in the operating energy range.}
\label{fig:rigel_linearity}
\end{figure}

Using an external test-pulse to inject a fixed charge at the CSA input, the linearity and full-scale range of the Readout Pixel Cell have been measured. The results are illustrated in Fig.~\ref{fig:rigel_linearity}, showing a non-linearity error, measured on the input energy range from 3 keV to 30 keV, between $\pm$ 2.7\%.

\subsection{Pulse Discrimination Threshold}
\begin{figure}[t]
\centering
\includegraphics[width=0.8\columnwidth]{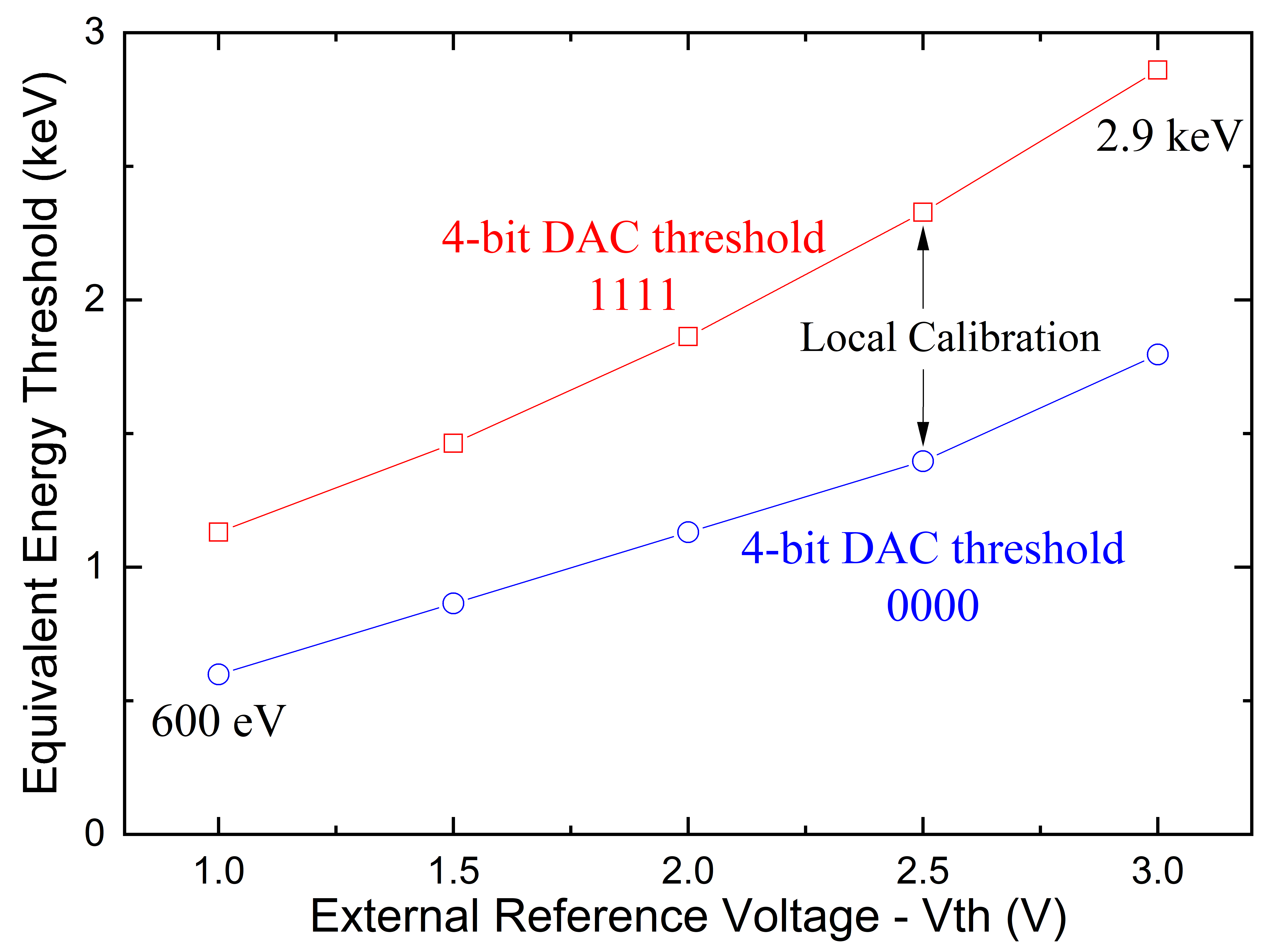}
\caption{Energy threshold experimental calibration of the RIGEL full-analog readout pixel cell (RPC). A global voltage reference (horizontal axis), acting as a coarse energy threshold, is generated by a top-level DAC, and distributed among all the RPCs. Each RPC can then internally adjust the local threshold using a local 4-bit DAC, performing a local fine tuning from code 0000b (blue curve) to 1111b (red curve).}
\label{fig:rigel_calibration}
\end{figure}

Experimental characterization was also done in stretch mode, to verify the correct functionality of the pulse discriminator and trigger logic. Since the internal trigger depends on the noise threshold of the amplitude discriminator, the noise thresholds have been measured and characterized, as well. At top-level, a coarse reference voltage is generated by a primary DAC to establish a global threshold level. At pixel level the noise threshold can be corrected with a fine threshold through an internal 4–bit DAC. In Fig.~\ref{fig:rigel_calibration} the experimental energy threshold obtained on the analog-RPC with an external reference voltage (simulating the global-threshold DAC), shows that the noise threshold can range approximately from 600~eV to 3~keV equivalent energy.

\section{Conclusions}
The architecture and functional specifications of the RIGEL ASIC, designed for the sparse readout a Silicon Pixel Drift Detector (PixDD) for space applications have been presented. Despite a low power budget ($\leq$ 0.55 mW/Channel on the analog section) and a strong constraint on the area occupation (300 \textmu m $\times$ 300 \textmu m), the RIGEL ASIC will be capable of integrating the complete signal processing chain, from the charge readout up to the digital signal processing. While the first flip-chip integration tests between the full PixDD matrix, and the multi-channel version of the RIGEL ASIC are still in progress, in the preset work we presented the complete characterization of the RPC analog front-end, which set the ultimate limit in terms of performance for the final multi-channel ASIC. Specifically, an optimum FWHM of 167 eV on the 5.9~keV line has been have been measured at 0\textdegree C, a linearity error below  $\pm$2.7\% has been measured over the 3~keV -- 30~keV input energy range, with an adjustable noise threshold ranging from 600~eV to 3~keV.

\section{Acknowledgements}
This work has been carried out in the framework of the ReDSoX and ADAM collaborations. We acknowledge
support from ASI-INAF Accordo Attuativo ADAM n. 2018-11-HH.0.

 \bibliographystyle{elsarticle-num} 
 \bibliography{refs}

\end{document}